\documentclass[journal=jpclcd,manuscript=letter]{achemso}

\usepackage{graphicx}
\usepackage{dcolumn}
\usepackage{bm}
\usepackage{lipsum}
\usepackage{lineno}
\usepackage[table,xcdraw]{xcolor}
\usepackage{multirow}
\usepackage{soul}
\usepackage{simplewick}
\usepackage{chemformula}
\usepackage{editing} 
\usepackage{hhline}
\usepackage{threeparttable}
\usepackage{xfrac}
\usepackage{multirow}
\usepackage{braket}
\usepackage[capitalize]{cleveref}

\newcommand{\ahat}{\hat{a}}
\newcommand{\bvec}[1]{\bm{\mathrm{#1}}}

\title{Towards accurate spin-orbit splittings from relativistic multireference electronic structure theory}

\author{Zijun Zhao}
\email{zijun.zhao@emory.edu}
\author{Francesco A. Evangelista}
\email{francesco.evangelista@emory.edu}
\affiliation{Department of Chemistry and Cherry Emerson Center for Scientific Computation, Emory University, Atlanta, Georgia 30322, USA}

\date{\today}

\begin{document}
\begin{tocentry}
\includegraphics[width=\textwidth]{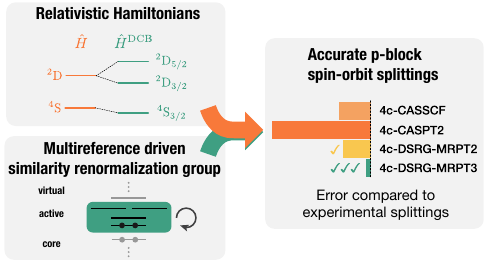}
\end{tocentry}

\begin{abstract}
Most nonrelativistic electron correlation methods can be adapted to account for relativistic effects, as long as the relativistic molecular spinor integrals are available, from either a four-, two-, or one-component mean-field calculation. 
However, relativistic multireference correlation methods remain a relatively unexplored area, with mixed evidence regarding the improvements brought by perturbative treatments. 
We report, for the first time, the implementation of state-averaged four-component multireference perturbation theories to second and third order based on the driven similarity renormalization group (DSRG). 
With our methods, named 4c-SA-DSRG-MRPT2 and 3, we find that the dynamical correlation included on top of 4c-CASSCF references can significantly improve the spin-orbit splittings in p-block elements and potential energy surfaces when compared to 4c-CASSCF and 4c-CASPT2 results. 
We further show that 4c-DSRG-MRPT2 and 3 are applicable to these systems over a wide range of the flow parameter, with systematic improvement from second to third order in terms of both improved error statistics and reduced sensitivity with respect to the flow parameter.
\end{abstract}

The accurate description of the electronic structure of systems containing heavy atoms requires a balanced treatment of both strong electron correlation and relativistic effects.\cite{vilkasRelativisticMultireferenceManybody1999,abeRelativisticCompleteActivespace2006} 
The former arises from the commonplace occurrence of near-degenerate configurations in such systems, and the latter arises from the $Z^2$ dependence of relativistic contribution to the the valence electron energy,\cite{pyykkoRelativisticEffectsChemistry2012} where $Z$ is the atomic number. 
The effects of spin-orbit coupling (SOC), which are completely absent from the nonrelativistic electronic Hamiltonian, are reflected in spectroscopic measurements,\cite{cunha202210.1021/acs.jpclett.2c00578,moitra202310.1021/acs.jpclett.2c03599,carbery201710.1021/acs.jpclett.7b00130} have an outsized influence on reaction energetics,\cite{czakoSurprisingQuenchingSpin2014,agarwal201110.1021/jz201124j} NMR shielding constants,\cite{malkin201310.1021/jz302146m} and play an important role in photophysics and photochemistry.\cite{fantacci201410.1021/jz402544r,zhang202010.1021/acs.jpclett.0c02135,valentine202210.1021/acs.jpclett.2c00207} 
Four-component (4c) mean-field and electron correlation methods employ Hamiltonians that explicitly couple the spin and orbital angular momentum degrees of freedom, and are the methods of choice to treat SOC effects in an \textit{ab initio} manner. 
However, relativistic electron correlation methods are not always able to consistently improve the description of SOC effects from relativistic mean-field references.\cite{zhangSpinOrbitCoupling2018}

Computational methods for treating strongly correlated systems have been an area of intense study.\cite{cottonTruncatedDavidsonMethod2022a,liuICIIterativeCI2016,schriberCommunicationAdaptiveConfiguration2016,sharmaSemistochasticHeatBathConfiguration2017,zhangIterativeConfigurationInteraction2020,greeneImprovedFastRandomized2020,leeTwentyYearsAuxiliaryField2022,boothFermionMonteCarlo2009,clelandCommunicationsSurvivalFittest2010,thomStochasticCoupledCluster2010,umrigarDiffusionMonteCarlo1993,morales-silvaFrontiersStochasticElectronic2021,needsContinuumVariationalDiffusion2009,zhangQuantumMonteCarlo2003,schollwockDensitymatrixRenormalizationGroup2005,whiteDensitymatrixAlgorithmsQuantum1993,whiteDensityMatrixFormulation1992,woutersDensityMatrixRenormalization2014,lyakhMultireferenceNatureChemistry2012,evangelistaPerspectiveMultireferenceCoupled2018}
Most nonrelativistic correlation methods can be readily adapted to account for relativistic effects within the no-pair approximation, as long as the complex-valued molecular spinor integrals are available,\cite{saueRelativisticHamiltoniansChemistry2011,wangClosedshellCoupledclusterTheory2008,dyallIntroductionRelativisticQuantum2007} after either a two- or four-component mean-field calculation. 
As such, considerable work has been devoted to developing mean-field methods. 
On the four-component side, the Dirac--Hartree--Fock (DHF, or 4c-HF) and complete active space self-consistent field (4c-CASSCF) \cite{reynoldsLargescaleRelativisticComplete2018,batesFullyRelativisticComplete2015,thyssenDirectRelativisticFourcomponent2008,jo/rgenaa.jensenRelativisticFourComponent1996} can be routinely carried out in several packages.\cite{saueDIRACCodeRelativistic2020,shiozakiBAGELBrilliantlyAdvanced2018,williams-youngChronusQuantumSoftware2020,sunPySCFPythonbasedSimulations2018,sunRecentDevelopmentsPySCF2020}
Depending on whether a 2c/4c-HF or CASSCF reference is used, these methods are further grouped into single-\cite{laerdahlDirectRelativisticMP21997,wangClosedshellCoupledclusterTheory2008} or multireference.\cite{fleigGeneralizedActiveSpace2001,watanabeFourcomponentRelativisticConfigurationinteraction2002,andersonFourcomponentFullConfiguration2020a,zhangSOiCIICISOCombining2022,wangRelativisticSemistochasticHeatBath2023}
Multireference relativistic methods have received proportionately more attention than their nonrelativistic cousins, as many systems where a relativistic study is warranted, such as late-row transition metal, lanthanide, and actinide complexes, exhibit strong configuration mixing in the ground state, and as a result require the zeroth order wave function to be multiconfigurational. 
Examples of relativistic multireference theories include Fock-space multireference coupled cluster (FSMRCC),\cite{vilkasRelativisticMultireferenceManybody1999} 4c internally contracted MRCI (ic-MRCI), CASPT2,\cite{abeRelativisticCompleteActivespace2006} NEVPT2,\cite{shiozakiRelativisticInternallyContracted2015} generalized van Vleck PT2 (GVVPT2),\cite{tamukongRelativisticGVVPT2Multireference2014} multireference M{\o}ller--Plesset (MRMP).\cite{vilkasRelativisticMultireferenceManybody1999,luExactTwoComponentRelativisticMultireference2022} 
In several ways, these methods reflect one or more limitations of their nonrelativistic counterparts: 1) lack of proper scaling with system size, 2) difficulties with scaling to large active spaces, 3) numerical divergences arising from the intruder state problem, and 4) insufficient accuracy in the description of dynamical electron correlation.

To address the abovementioned insufficiencies of relativistic multireference theories, we have explored a 4c generalization of the multi-reference driven similarity renormalization group (MR-DSRG) approach.\cite{evangelistaDrivenSimilarityRenormalization2014,liMultireferenceTheoriesElectron2019} 
The MR-DSRG is a size-extensive, low-scaling, numerically robust, and systematically improvable many-body formalism for treating dynamical electron correlation starting from strongly correlated reference states.\cite{evangelistaDrivenSimilarityRenormalization2014,liMultireferenceTheoriesElectron2019} 
In this paper, we show that a combination of 4c strongly correlated reference states and the MR-DSRG is a very promising route to systematically account for relativistic and correlation effects.
We demonstrate this point by reporting the first implementation of 4c MR-DSRG methods truncated to second- and third-order in perturbation theory (4c-DSRG-MRPT2/3) and benchmarking them on the spin-orbit splittings of second- to fourth-row atoms, which previous work has shown to be a weak point for four-component correlation methods.\cite{zhangSpinOrbitCoupling2018}

The starting point for formulating the 4c-MR-DSRG is first-quantized quasi-relativistic many-electron Hamiltonian:\cite{reiherRelativisticQuantumChemistry2014}
\begin{equation}
\label{eq:fqhamil}
    H=\sum_{i}^Nh_{\mathrm{D}}(i)+\frac{1}{2}\sum_{i\neq j}^Ng(i,j)+V_{\mathrm{NN}},
\end{equation}
where $h_{\mathrm{D}}(i)$ is the one-electron Dirac Hamiltonian for electron $i$, and $g(i,j)$ is the two-electron interaction operator for an electron pair $i$ and $j$. 
The latter can contain either the bare-Coulomb operator, leading to the Dirac--Coulomb (DC) Hamiltonian, or can additionally include the Gaunt and gauge terms, leading to the Dirac--Coulomb--Gaunt (DCG) or Dirac--Coulomb--Breit (DCB) Hamiltonians.
These operators are defined as
\begin{align}
    g^{\mathrm{DCB}}(i,j) = \underbrace{\vphantom{+\frac{(\bvec{\alpha}_i\cdot\bvec{\nabla}_i)(\bvec{\alpha}_j\cdot\bvec{\nabla}_j)}{2r_{ij}^3}}\frac{1}{r_{ij}}}_{\mathrm{Coulomb}}
    \underbrace{\vphantom{+\frac{(\bvec{\alpha}_i\cdot\bvec{\nabla}_i)(\bvec{\alpha}_j\cdot\bvec{\nabla}_j)}{2r_{ij}^3}}-\frac{\bvec{\alpha}_i\cdot\bvec{\alpha}_j}{r_{ij}}}_{\mathrm{Gaunt}}
    \underbrace{+\frac{(\bvec{\alpha}_i\cdot\bvec{\nabla}_i)(\bvec{\alpha}_j\cdot\bvec{\nabla}_j)}{2r_{ij}^3}}_{\mathrm{gauge}},
\end{align}
where $\bvec{\alpha}_i$ is a $3$-vector of the Dirac matrices for particle $i$.
The Coulomb operator contains charge-charge and spin-same-orbit interactions, the Gaunt and Breit operators contain additional current-current and spin-other-orbit interactions, and a gauge term resulting from working in the Coulomb gauge. \cite{saueRelativisticHamiltoniansChemistry2011,dyallIntroductionRelativisticQuantum2007}
Following common practice, we use the DCB Hamiltonian in combination with the no-pair approximation, whereby the many-body basis is constructed from positive-energy spinors only.\cite{reiherRelativisticQuantumChemistry2014}
The four-component molecular spinors $\{\psi_i(\bm{r})\}$ are expanded in a two-spinor basis set $\{\phi_{\mu}^{S/L}\}$ for the small/large components.
To avoid variational collapse, we impose the kinetic balance condition, $2c\phi_{\mu}^S=\bm{\sigma}\cdot\bm{p}\phi_{\mu}^L$, where $\bvec{\sigma}$ is a $3$-vector of Pauli matrices.\cite{stantonKineticBalancePartial1984a}
The large component 2-spinor basis functions comprise standard scalar Gaussian basis functions multiplied by 2-spinor spherical basis functions.\cite{reiherRelativisticQuantumChemistry2014}

We consider the most general, state-averaged formulation of the MR-DSRG approach, which can target one or multiple electronic states in one computation.
We approximate the $k$-th electronic state to zeroth-order with 4c-CASSCF states of the form
\begin{equation}
    |\Psi^{(k)}\rangle=\sum_{\mu=1}^{N_{\mathrm{CAS}}} C_{\mu}^{(k)}|\Phi^{\mu}\rangle,
\end{equation}
where $N_{\mathrm{CAS}}$ is the number of CAS determinants. 
In 4c-CASSCF, the spinor orbitals $\{\psi_i\}$ that enter into Slater determinants $\{\Phi^{\mu}\}$ and the complex CASCI coefficients $C_{\mu}^{(k)}$ are simultaneously optimized \textit{via} the minimax principle.\cite{talmanMinimaxPrincipleDirac1986} 
No positronic orbitals enter into the determinants, in what is known as the ``no dressed pair'' or ``no virtual pair'' approximation (NVPA).\cite{jo/rgenaa.jensenRelativisticFourComponent1996,hoyerCorrelatedDiracCoulomb2023a} 
Furthermore, to ensure the correct degeneracies of the electronic states, we use the state-averaged CASSCF formalism (4c-SA-CASSCF)\cite{wernerQuadraticallyConvergentMCSCF1981}, where the average energy of a chosen set of states is optimized with respect to the variational parameters. \cite{helgakerMolecularElectronicStructure2000}

\begin{figure}[!htp]
    \centering
    \includegraphics[width=5in]{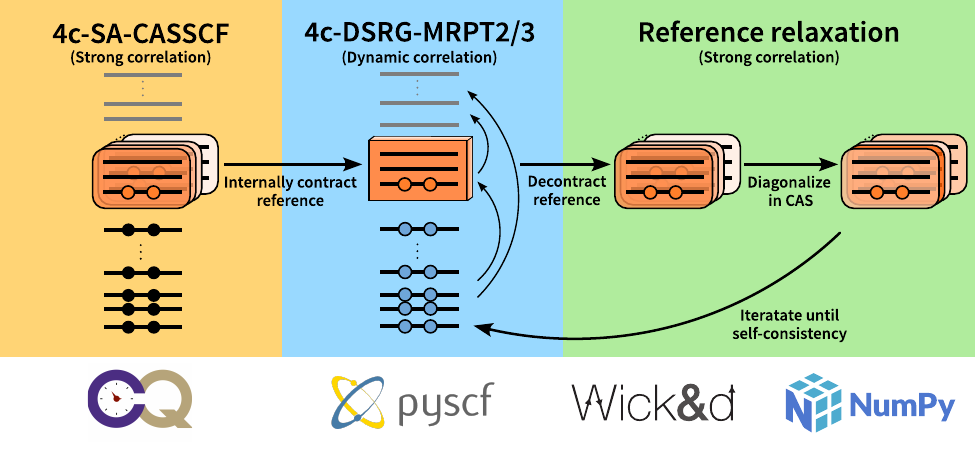}
    \caption{Summary of the 4c-SA-DSRG-MRPT2/3 algorithm. The iterative reference relaxation process is terminated once a threshold is reached.}
    \label{fig:summary}
\end{figure}

\begin{figure}[!htp]
    \centering
    \includegraphics[width=3.375in]{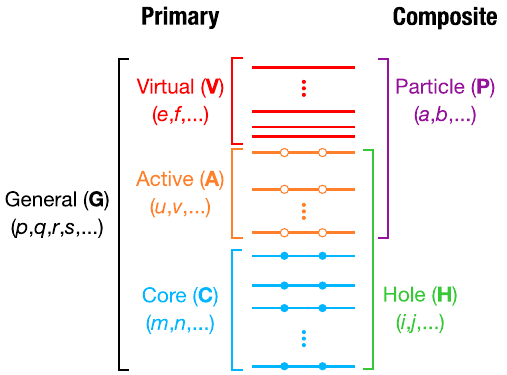}
    \caption{The partitioning of the orbital basis, and the orbital indices used in the MR-DSRG formalism.}
    \label{fig:orb_part}
\end{figure}

To account for dynamical electron correlation, we combine 4c-(SA-)CASSCF wave functions with the multireference driven similarity renormalization group (MR-DSRG) formalism.\cite{evangelistaDrivenSimilarityRenormalization2014}
The starting point of the 4c-MR-DSRG formalism is the second-quantized version of the quasi-relativistic no-pair Hamiltonian:
\begin{equation}
    \hat{H} = E_0+\sum^{\mathbf{G}}_{pq}f^q_p\{\ahat^p_q\}+\frac{1}{4}\sum^{\mathbf{G}}_{pqrs}v^{rs}_{pq}\{\ahat^{pq}_{rs} \}.
\end{equation}
The orbital indexing convention and definition of orbital spaces used in this work reflects the partitioning of the orbitals into core, active, and virtual sets, as shown in \cref{fig:orb_part}.
The quantities $f^q_p$ that enter in the Hamiltonian are elements of the generalized Fock matrix, $f^q_p = h^q_p+\sum_{ij}^{\mathbf{H}}v_{pi}^{qj}\gamma^i_j$, where $\gamma^i_j$ is the one-body density matrix defined for a general state $\Psi$ as $\gamma^i_j = \braket{\Psi | \hat{a}^\dagger_i \hat{a}_j | \Psi}$.
The MR-DSRG performs a continuous unitary transformation of the Hamiltonian:
\begin{equation}
    \hat{H} \mapsto \bar{H}=e^{-\hat{A}}\hat{H}e^{\hat{A}}.
\end{equation}
The operator $\hat{A}$ is anti-Hermitian, and is expressed in terms of a cluster operator as $\hat{A}=\hat{T}-\hat{T}^{\dagger}$.
$\hat{T}$ is parameterized as in the internally-contracted generalization of coupled cluster theory\cite{evangelistaOrbitalinvariantInternallyContracted2011,hanauerPilotApplicationsInternally2011,evangelistaSequentialTransformationApproach2012} as $\hat{T} = \hat{T}_1 + \ldots \hat{T}_n$, where a general $k$-body term is expressed as
\begin{equation}
	\hat{T}_k = \frac{1}{({k!})^2} \sum_{ij\cdots}^\mathbf{H} \sum_{ab\cdots}^\mathbf{P} t_{ab\cdots}^{ij\cdots} \{ \hat{a}_a^\dagger \hat{a}_b^\dagger \cdots \hat{a}_j \hat{a}_i \}\equiv \frac{1}{({k!})^2} \sum_{ij\cdots}^\mathbf{H} \sum_{ab\cdots}^\mathbf{P} t_{ab\cdots}^{ij\cdots}\{\hat{a}^{ab\dots}_{ij\dots}\},
\end{equation}
where $\{\cdot\}$ indicates normal-ordering with respect to a correlated reference state $\Psi$ or ensemble density.\cite{kutzelniggNormalOrderExtended1997} 

\begin{figure}[!hbt]
\center
  \includegraphics[width=6.5in]{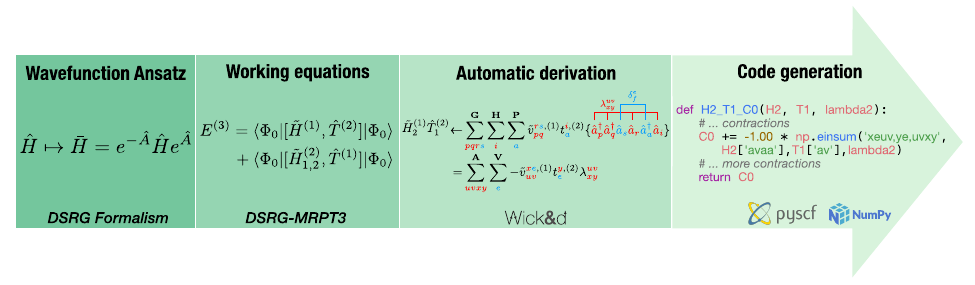}
  \caption{Overview of the automated strategy used in this paper to implement the 4c DSRG methods. Starting from the left, the DSRG ansatz is perturbatively expanded to give working equations for the energy corrections. These are then turned into tensor contractions using the open-source \textsc{Wick\&d} package.\cite{evangelistaAutomaticDerivationManybody2022} Contractions generated by \textsc{Wick\&d} are then automatically compiled into executable Python code.}
  \label{fig:codegen}
\end{figure}

The effect of the MR-DSRG transformation is to bring the Hamiltonian into a more band-diagonal form by zeroing the couplings between the multi-configurational reference $\ket{\Psi_0}$ and the excited configurations, \textit{i.e.}, the set of internally-contracted determinants $\ket{\Psi_{ij\dots}^{ab\dots}}=\{\hat{a}^{ab\dots}_{ij\dots}\}\ket{\Psi_0}$.\cite{mukherjeeNormalOrderingWicklike1997,kutzelniggNormalOrderExtended1997} 
To avoid numerical divergences caused by intruder states, in the DSRG we solve a regularized equation that enforces the decoupling of the off-diagonal elements of $\bar{H}$, denoted as $[\bar{H}]_{\mathrm{od}}$. This equation reads as $[\bar{H}(s)]_{\mathrm{od}}=\hat{R}(s)$, where $\hat{R}(s)$ is a regularization term derived by a second-order perturbative analysis of the flow similarity renormalization group.\cite{evangelistaDrivenSimilarityRenormalization2014,wegnerFlowequationsHamiltonians1994,kehreinFlowEquationApproach2006a}
Several authors have pointed out the benefits of introducing a parameterized denominator shift or a regularizer of the first-order amplitudes in second-order perturbation theory---both in its single and multireference versions.\cite{roosMulticonfigurationalPerturbationTheory1995,sheeRegularizedSecondOrderMoller2021}
In the DSRG, this role is played by the flow parameter $s$, which in typical applications is found to be optimal for values in the range $s\in[0.5,2]$, depending on the truncation level.\cite{wangAssessmentStateAveragedDriven2023}
To generate equations and corresponding tensor contractions for the 4c relativistic extension of the MR-DSRG truncated to second- and third-order in perturbation theory, we have developed the automated code implementation pipeline depicted in \cref{fig:codegen}.
This approach enables the rapid implementation of highly complex electronic structure theories, removing the potential for human error.
This workflow uses our code generator \textsc{Wick\&d}\cite{evangelistaAutomaticDerivationManybody2022} coupled to relativistic integrals obtained from \textsc{PySCF}\cite{sunLibcintEfficientGeneral2015,sunPySCFPythonbasedSimulations2018,sunRecentDevelopmentsPySCF2020}.
Since 4c-CASSCF is currently not available in \textsc{PySCF}, we generated the corresponding spinor coefficients using the \textsc{ChronusQ} package.\cite{williams-youngChronusQuantumSoftware2020}

\begin{figure*}[!htp]
    \centering
    \includegraphics[width=6.5in]{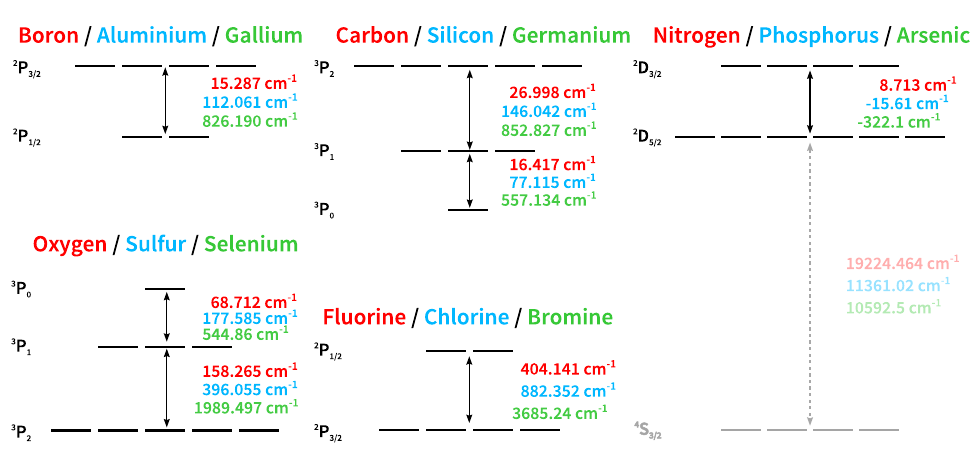}
    \caption{A scheme of the spin-orbit splittings of the p-block elements discussed. State-averaging was performed over all states shown. The greyed-out splittings are excluded from analysis and are only included in the state-averaging procedure to ensure variational stability. A switch between the $^2$D$_{3/2}$ and $^2$D$_{5/2}$ states in the group 15 elements is indicated by negative splittings. Energy levels accessed from the NIST Atomic Spectra Database.\cite{kramidaNISTAtomicSpectra1999}}
    \label{fig:pblock}
\end{figure*}
%

\begin{figure}[!htp]
    \centering
    \includegraphics[width=3.375in]{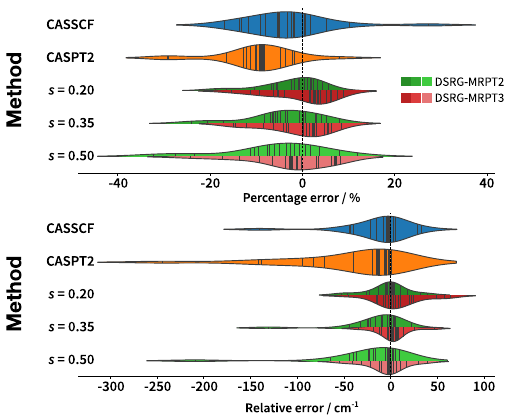}
    \caption{Distributions of the errors in the computed spin-orbit splittings. The dashed vertical lines at $0$ represent experimental splittings. The plots were generated with the \texttt{violinplot} function of \textsc{seaborn},\cite{seaborn,matplotlib} where the data points and the kernel density estimates of the splitting errors of each method are shown. The outlier elements of each plot are labeled.}
    \label{fig:violin}
\end{figure}

Computational details are found in Section 1 of the Supplementary Information.
To examine the ability of the 4c-DSRG-MRPT2/3 to improve systematically upon 4c-CASSCF, we computed spin-orbit splittings of second- to fourth-row p-block elements, and the spectroscopic constants for the hydroxyl radical.
The p-block splittings span three orders of magnitude, from a few reciprocal centimeters (cm$^{-1}$) to a few thousand, as shown in \cref{fig:pblock}. 
Therefore, they form a good set of benchmarks for both the absolute and relative accuracies of relativistic methods.

In \cref{fig:violin}, we compare error statistics for the splittings for the 15 p-block elements calculated with 4c-CASSCF, 4c-SA-DSRG-MRPT2/3, 4c-CASPT2, and 4c-MR-CISD+Q (taken from the work by Zhang \textit{et al}.\cite{zhangSpinOrbitCoupling2018}) to the experimental splittings for three values of the flow parameter (0.2, 0.35, and 0.5 $E_{\mathrm{h}}^{-2}$), which cover the range of flow parameters used in non-relativistic computations and that produce results competitive with 4c CASSCF and CASPT2.
Table S1 reports splittings for each state with optimal flow parameters that achieve the smallest mean absolute error in cm$^{-1}$ for the whole set of elements for each of the methods respectively ($0.24$ $E_{\mathrm{h}}^{-2}$ for MRPT2 and $0.35$ $E_{\mathrm{h}}^{-2}$ for MRPT3). 
Looking at the distribution of errors (top plot in \cref{fig:violin}), both MRPT2 and 3 achieve better average performance on the test set of atoms than CASSCF and the two other multireference electron correlation methods (except for $s = 0.5$ $ E_{\mathrm{h}}^{-2}$ in MRPT2).
Improving the treatment of dynamical electron correlation from second to third order results in a systematic reduction of the error in the splittings and a decrease in flow parameter sensitivity.
Across all methods, the main outliers in percentage error (bottom plot in \cref{fig:violin}) are elements with small (10-150 cm$^{-1}$) splittings, with the exceptions of arsenic (322 cm$^{-1}$).
The complete dataset for the splittings can be found in Sections 2 and 3 in the Supplementary Information.
When compared according to signed errors, irrespective of the value of $s$, the main outliers are fourth-row elements with large splittings, with selenium and gallium being the worst offenders.
We also note that DSRG-MRPT3 achieves sub-wavenumber accuracy for boron and carbon, and rather impressively, less than 2 cm$^{-1}$ errors for the fourth-row selenium and bromine.
4c-CASPT2 and 4c-MR-CISD+Q can be seen to consistently and severely underestimate spin-orbit splittings. 
This behavior was also pointed out in Zhang \textit{et al.}\cite{zhangSpinOrbitCoupling2018} to occur for smaller and larger basis sets than the ones used here.
In Table S2, we report additional splittings represented by dashed arrows in \cref{fig:pblock}.
We again observe that MRPT2/3 outperform CASPT2, which in turn outperforms CASSCF in this set of transitions. This shows that the 4c-SA-DSRG formalism is capable of accurately capturing SOC effects in higher excited states.

\begin{figure}[!htp]
    \centering
    \includegraphics[width=3.375in]{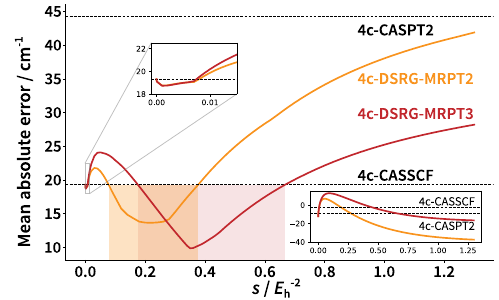}
    \caption{Variation of the mean average error of 4c-DSRG-MRPT2 and MRPT3 as a function of the flow parameter, $s$, for all of the splittings computed using the DCB Hamiltonian with DCB CASSCF molecular spinor coefficients. The main curves show absolute errors, and the bottom insets show signed errors. The middle insets zoom into the region close to $s=0$.}
    \label{fig:dcb-curve}
\end{figure}

After assessing the quality of DSRG-based methods, we turn to the question of the dependence of our results on the flow parameter and its optimal choice for spin-orbit coupling computations.
To this end, we studied the variation of the mean absolute percentage error in the splittings as the flow parameter is varied in the range of $s\in[0,1.3]$ $E_{\mathrm{h}}^{-2}$.
This quantity is plotted in \cref{fig:dcb-curve} for both the DSRG-MRPT2/3.
In the middle inset, we first observe the correct limiting behavior for $s=0$, where  MRPT2 and MRPT3 reduce to CASSCF.
For both PT2 and PT3, the mean absolute percentage errors are smaller than the CASSCF  and CASPT2 values for a large range of $s$, up to $s\approx0.75$ $E_{\mathrm{h}}^{-2}$ for PT3. 
For PT2, and PT3 especially, these values overlap with the commonly used range $s\in[0.35,1.0]$. 
The kinks in the main curves are artifacts of the absolute values, as the mean signed error curves (bottom insets) are smooth.
These results show that the \textit{error profiles} of MRPT2 and MRPT3 do not simply interpolate between those of CASSCF and CASPT2, and bring systematic improvements from a balanced treatment of dynamical correlation.

Sections 5 and 6 in the Supplementary Information report summary error statistics broken down into groups and periods.
An analysis of this data shows that second-row elements display significantly larger absolute and percentage errors, which cannot be simply attributed to the elements having smaller splittings. 
This is most likely due to the poor description of the core region by the relatively small uncontracted cc-pVTZ basis set for boron through fluorine. 
As SOC effects decay as $1/r^3$,\cite{atkinsMolecularQuantumMechanics2010} having variational flexibility in the core region is crucial: Visscher and Dyall\cite{visscherRelativisticCorrelationEffects1996} found that the cc-pVTZ basis augmented with an additional tight p function reduced the error in the spin-orbit splitting of \ch{F2} tenfold to $1\%$. 
We have tested this hypothesis by following Visscher and Dyall and added a tight p function with the exponent of 250.83491 to the cc-pVTZ basis of fluorine. 
The resulting $^2$P$_{\sfrac{3}{2}}\rightarrow^2$P$_{\sfrac{1}{2}}$ splitting error of fluorine for MRPT3 using $s=0.35$ $E_{\mathrm{h}}^{-2}$ improved from $-12.46$ cm$^{-1}$ ($-3.1$\%) to $-1.04$ cm$^{-1}$ ($-0.3$\%).
Analogous data for the remaining second-row elements is reported in Section 8 of the Supplementary Information.
From these results, we can see that except for the N $^2$D$_{\sfrac{5}{2}}\rightarrow^2$D$_{\sfrac{3}{2}}$ splitting, the provision of more core flexibility significantly improves the description of second-row valence SOC effects with a negligible increase in basis size. 
The improvement in the underlying CASSCF splittings largely drives the improved description. 
However, DSRG-MRPT2 and 3 still provide an improvement over the CASSCF results.
Another aspect we investigated is a reduction of the cost of the 4c-CASSCF procedure, which in our computations, especially on third- and fourth-row elements, is overwhelmingly the most time-consuming step.
In Section 7 of the Supplementary Information, we report additional data for a mixed scheme aimed at reducing this cost that uses a combination of the DC and DCB Hamiltonians in the 4c-CASSCF and DSRG-MRPT2 computations.
The data presented therein supports the use of such a mixed scheme, as the reduction in accuracy is negligible.

Lastly, we consider the accuracy of potential energy surfaces computed with the 4c-DSRG-MRPT methods.
For example, in \cref{tab:oh} we report the spectroscopic constants of the ground state of the hydroxyl radical computed with 4c-CASSCF and 4c-DSRG-MRPT3.
In this molecule, SOC results in a zero-field splitting (ZFS) of the ground $\mathrm{X}\ ^2\Pi$ state into two states with $\Omega=3/2$ and $1/2$ respectively.\cite{gui-xia200610.1088/1009-1963/15/5/022,muck201210.1063/1.3694132,maeda201510.1088/1367-2630/17/4/045014} 
Our results show that although both  4c-CASSCF and 4c-DSRG-MRPT predict ZFSs in excellent agreement with the experimental value,\cite{liu202010.1186/s13321-020-00433-8} the spectroscopic constant from the 4c-DSRG-MRPT3 potential display significantly smaller errors.
This point is particularly evident for the harmonic vibrational frequency of OH, which 4c-CASSCF underestimates by 210.9 cm$^{-1}$ vs. 8.7 cm$^{-1}$ for the 4c-DSRG-MRPT3.
Finally, we computed the potential energy surfaces of the $\mathrm{X}\ ^2\Pi_{3/2}$ and $^2\Pi_{1/2}$ states.

\begin{table}
\renewcommand{\arraystretch}{1.5}
\begin{tabular}{lrrrrr}
	\hhline{======}
	Method & $\Delta r_e$ (pm) & $\Delta \omega_e$ (cm$^{-1}$) & $\Delta \omega_e x_e$ (cm$^{-1}$) & $\Delta D_0$ (eV) & $\Delta$ZFS (cm$^{-1}$)\\ \hline
	4c-SA-CASSCF& $ 0.337$ & $-210.9$ & $30.2$ & $-0.914$ & $-5.2$ \\
	4c-SA-PT2   & $ 0.472$ & $ -71.3$ &$-20.5$ & $-0.174$ & $-6.7$ \\
	4c-SA-PT3   & $-0.112$ & $  -8.7$ & $10.1$ & $-0.224$ & $-4.0$ \\ \hline
    Exp.        & $96.966$ & $3737.8$ & $84.9$ & $ 4.392$ & $139.2$\\ \hhline{======}
\end{tabular}
\caption{Spectroscopic constants of the $\mathrm{X}\ ^{2}\Pi$ ground state of the OH radical calculated with different methods. Shown are the differences with respect to experimental values, taken from the Diatomic Molecular Spectroscopy Database.\cite{liu202010.1186/s13321-020-00433-8,huber.1979.10.1007/978-1-4757-0961-2} The (adiabatic) zero-field splitting includes anharmonic zero-point vibrational energy contributions.}
\label{tab:oh}
\end{table}


In summary, we have presented second- and third-order four-component multi-reference perturbation theories based on the driven similarity renormalization group formalism, 4c-DSRG-MRPT2/3. We benchmarked these methods on the spin-orbit splittings of second- to fourth-row p-block atoms, showing that they outperform both the underlying 4c-CASSCF and other four-component electron correlation methods, namely 4c-MR-CISD+Q and 4c-CASPT2. 
We have further shown that 4c-DSRG-MRPT2 and 3 are applicable to these systems over a wide range of the flow parameter, with systematic improvements in error metrics and sensitivity with respect to $s$ from second to third order.
In our calculations, most of the wall time is spent in the integral transformation step, which is known to be the main drawback of four-component theories. 
Exact two-component (X2C) theories with atomic mean-field spin-orbit effects (amfX2C)\cite{sikkemaMolecularMeanfieldApproach2009,liuAtomicMeanfieldSpinorbit2018,zhangAtomicMeanFieldApproach2022,knechtExactTwocomponentHamiltonians2022} are an attractive way to reduce the computational effort not only of integral transformations but also in the SCF iterations themselves, depending on the flavor of X2C used.
In the future, we plan to explore the combination of amfX2C with two-component MR-DSRG methods to extend multireference relativistic computations to larger systems.
Overall, these preliminary results show that a combination of relativistic Hamiltonians and multireference DSRG methods could open many new avenues in modeling the chemistry of open-shell species containing transition metals and heavy elements.


\section*{Acknowledgements}
This research was supported by the U.S. Department of Energy under Award DE-SC0024532 and a Camille Dreyfus Teacher-Scholar Award (No. TC-18-045).
\section*{Data availability}
The data that support the findings of this study are available within the article, its supplementary material, and from the corresponding author upon reasonable request.
Software that can reproduce the data presented within this work is available in an accompanying public GitHub repository.\cite{gitrepo}

\bibliography{main}
\end{document}



\title{Supplementary information: ``Towards accurate spin-orbit splittings from relativistic multireference electronic structure theory"}

\author{Zijun Zhao}
\email{zijun.zhao@emory.edu}
\author{Francesco A. Evangelista}
\email{francesco.evangelista@emory.edu}
\affiliation{Department of Chemistry and Cherry Emerson Center for Scientific Computation, Emory University, Atlanta, Georgia 30322, USA}

\date{\today}

\maketitle

\setcounter{table}{0}
\renewcommand{\thetable}{S\arabic{table}}%
\setcounter{figure}{0}
\renewcommand{\thefigure}{S\arabic{figure}}%

\section{Computational details}
We used the Dirac--Coulomb--Breit (DCB) Hamiltonian with the uncontracted Dunning's correlation-consistent polarized valence triple-zeta (uc-cc-pVTZ) basis set,\cite{dunningGaussianBasisSets1989,woonGaussianBasisSets1993,wilsonGaussianBasisSets1999} finite Gaussian nuclei,\cite{visscherDIRACFOCKATOMIC1997} freezing non-valence orbitals in all third- and fourth-row atoms from the correlated calculations.
 
For the p-block elements, whenever the neutral atom cannot be converged by the 4c-HF procedure, the corresponding ion with the configuration of the closest noble gas is used instead (\textit{e.g.}, \ch{Al^3+}, \ch{Se^2-}), and the molecular spinor coefficients are used as initial guesses for the 4c-CASSCF calculations. 
State averaging is performed on all states depicted in \cref{fig:pblock} with equal weights for each state, with the sum of weights normalized to unity. 
The reason to include additional states in the state averaging procedure is to avoid root-flipping problems, or to avoid biasing the CASSCF convergence (in the case of including the $^4$S$_{\sfrac{3}{2}}$ states for group 15 elements). 
The states shown correspond to the lowest 6, 10, 14, 10, 6 states of the 4c-SA-CASSCF solutions for groups 13 to 17 respectively.
In our DSRG calculations, full (iterative) reference relaxation is performed for all calculations.

For the calculations involving the hydroxyl radical, a complete active space of 5 electrons in 8 spinor orbitals is used, which corresponds to the $2p_{\sigma}$, the doubly degenerate $2p_{\pi}$, and the $2p_{\sigma}^*$ MOs.
The state-averaged calculations averages over the lowest 6 states with equal weights, corresponding to the doubly degenerate $\mathrm{X}\ ^2\Pi_{3/2}$, $^2\Pi_{1/2}$, and $\mathrm{
A}\ ^2\Sigma^+$ states. 
The default convergence criteria from \textsc{Chronus Quantum} are used.
For all DSRG-MRPT2/3 computations, a flow parameter of 0.5 $E_{\mathrm{h}}^{-2}$ is used.
The spectroscopic constants are obtained from the \texttt{psi4.diatomic.anharmonicity} function provided by the open-source package \textsc{Psi4},\cite{smith.2020.10.1063/5.0006002} which is an implementation of the algorithm described by Bender \textit{et al}.\cite{bender.2014.10.1063/1.4862157}.
A grid size of 0.005 \AA\ is used in the range of  $[0.920,1.020]$ \AA, with a finer grid of 0.001 \AA\ in the range of $[0.960,0.990]$, for a total of 45 points, to determine the constants.
The dissociation energies, $D_0$, are calculated using state-specific CASSCF references, as state-averaged CASSCF calculations using the same settings are not feasible due to multiple degenerate states coming in from higher (untracked) excited states.\cite{easson.1973.10.1139/p73-068}

\section{The 15 `main' splittings computed with different methods}
\begin{table}[!htp]
\footnotesize
\begin{threeparttable}
\begin{tabular}{lcccccc}
\hhline{=======}
Splitting & Exp. & CASSCF & CASPT2 & MR-CIDS+Q & MRPT2 & MRPT3 \\
&&&&&($s=0.24$)&($s=0.35$)\\ \hline
B $^2$P$_{\sfrac{1}{2}}\rightarrow^2$P$_{\sfrac{3}{2}}$  & 15.29   & 13.25   & 14.25    & 13.91   & 13.99   & 14.32              \\
C $^3$P$_{0}\rightarrow^3$P$_{1}$                        & 16.42   & 14.94   & 16.96    & 15.40   & 14.93   & 17.20              \\
N $^2$D$_{\sfrac{5}{2}}\rightarrow^2$D$_{\sfrac{3}{2}}$  & 8.71    & 11.09   & 8.16     & 9.40\tnote{1}&9.41& 7.89                \\
O $^3$P$_{2}\rightarrow^3$P$_{1}$                        & 158.27  & 153.24  & 129.99   & 152.52  & 145.35  & 138.92            \\
F $^2$P$_{\sfrac{3}{2}}\rightarrow^2$P$_{\sfrac{1}{2}}$  & 404.14  & 382.58  & 380.46   & 388.38  & 384.70  & 391.68            \\
Al $^2$P$_{\sfrac{1}{2}}\rightarrow^2$P$_{\sfrac{3}{2}}$ & 112.06  & 96.81   & 106.60   & 106.96  & 106.70  & 108.35            \\
Si $^3$P$_{0}\rightarrow^3$P$_{1}$                       & 77.12   & 72.89   & 78.90    & 73.76   & 69.94   & 81.52              \\
P $^2$D$_{\sfrac{3}{2}}\rightarrow^2$D$_{\sfrac{5}{2}}$  & 15.61   & 15.34   & 14.01    &N/A\tnote{2}& 11.08   & 13.38              \\
S $^3$P$_{2}\rightarrow^3$P$_{1}$                        & 396.06  & 398.64  & 386.02   & 383.94  & 355.94  & 400.61            \\
Cl $^2$P$_{\sfrac{3}{2}}\rightarrow^2$P$_{\sfrac{1}{2}}$ & 882.35  & 886.86  & 894.86   & 861.80  & 867.69  & 888.44            \\
Ga $^2$P$_{\sfrac{1}{2}}\rightarrow^2$P$_{\sfrac{3}{2}}$ & 826.19  & 685.92  & 776.51   & 745.97  & 743.28  & 791.62            \\
Ge $^3$P$_{0}\rightarrow^3$P$_{1}$                       & 557.13  & 512.35  & 553.07   & 502.94\tnote{3}& 485.56  & 570.25            \\
As $^2$D$_{\sfrac{3}{2}}\rightarrow^2$D$_{\sfrac{5}{2}}$ & 322.10  & 354.53  & 324.93   &N/A\tnote{2}& 227.88  & 327.81            \\
Se $^3$P$_{2}\rightarrow^3$P$_{1}$                       & 1989.50 & 1949.63 & 1917.35  & 1900.34\tnote{3}& 1745.74 & 1991.36          \\
Br $^2$P$_{\sfrac{3}{2}}\rightarrow^2$P$_{\sfrac{1}{2}}$ & 3685.24 & 3683.62 & 3704.50  & 3540.14 & 3546.46 & 3683.90          \\ \hline
MAE &--&21.2&49.0& 33.4\tnote{4} &15.6&7.5\\ 
MAPE &--&7.81\%&10.7\%&5.60\%\tnote{4}&4.98\%&4.63\% \\
\hhline{=======}
\end{tabular}
\begin{tablenotes}
\scriptsize
\item [1] Two core spinors were frozen for nitrogen.
\item [2] The 4c-MR-CISD+Q computations were intractable for these atoms.
\item [3] The uncontracted cc-pVDZ basis set was used for these atoms.
\item [4] Unavailable data points have been omitted from averaging.
\end{tablenotes}
\end{threeparttable}
\caption{Comparison between the spin-orbit splittings of the 15 second- to fourth-row p-block elements calculated with 4c-SA-CASSCF, 4c-CASPT2, 4c-MR-CISD+Q, and MRPT2 and 3 to the experimental splittings. Flow parameters, $s$, are in units of $E_{\mathrm{h}}^{-2}$, and all results are reported in units of cm$^{-1}$, unless otherwise noted. ``Exp.'' stands for experimental splitting, ``MAE'' stands for mean absolute error, and ``MAPE'' stands for mean absolute percentage error.}
\label{tab:best-spl}
\end{table}

\section{The 9 additional splittings not included in summary statistics}
\begin{table}[!htp]
\footnotesize
\begin{threeparttable}
\begin{tabular}{lcccccc}
\hhline{=======}
Splitting & Exp. & CASSCF & CASPT2 & MR-CIDS+Q & MRPT2 & MRPT3 \\
&&&&&($s=0.24$)&($s=0.35$)\\ \hline
C $^3$P$_1\rightarrow^3$P$_2$  & 27.00    & 23.82    & 23.95    & 24.88    & 28.48     & 28.72     \\
N $^4$S$_{\sfrac{3}{2}}\rightarrow^2$D$_{\sfrac{5}{2}}$ & 19224.46 & 22910.28 & 20436.89 & 20095.77\tnote{1} & 20231.34  & 19952.54  \\
O $^3$P$_1\rightarrow^3$P$_0$ & 68.71    & 66.99    & 62.55    & 65.85    & 54.49     & 58.32     \\
Si $^3$P$_1\rightarrow^3$P$_2$ & 146.04   & 138.28   & 133.25   & 140.43\tnote{3}  & 149.20    & 154.03    \\
P $^4$S$_{\sfrac{3}{2}}\rightarrow^2$D$_{\sfrac{5}{2}}$ & 11361.02 & 15305.18 & 12561.03 & N/A\tnote{2}      & 13302.41  & 12666.93  \\
S $^3$P$_1\rightarrow^3$P$_0$ & 177.59   & 180.62   & 161.92   & 173.46   & 175.36    & 181.66    \\
Ge $^3$P$_1\rightarrow^3$P$_2$ & 852.83   & 809.65   & 711.39   & 814.20  & 857.09    & 878.13    \\
As $^4$S$_{\sfrac{3}{2}}\rightarrow^2$D$_{\sfrac{5}{2}}$& 10592.50 & 14289.54 & 11799.71 & N/A\tnote{2}      & 12567.66  & 12034.45  \\
Se $^3$P$_1\rightarrow^3$P$_0$& 544.86   & 572.88   & 532.84   & 566.74\tnote{3}  & 573.81    & 590.89  \\ \hline
MAPE &--&13.4\%&9.4\%&4.5\%\tnote{4}&8.5\%&7.7\% \\
\hhline{=======}
\end{tabular}
\begin{tablenotes}
\scriptsize
\item [1] Two core spinors were frozen for nitrogen.
\item [2] The 4c-MR-CISD+Q computations were intractable for these atoms.
\item [3] The uncontracted cc-pVDZ basis set was used for these atoms.
\item [4] As the most error-prone data points were intractable for MRCI, the average is not appropriate for comparison with other methods.
\end{tablenotes}
\end{threeparttable}
\caption{Comparison between additional spin-orbit splittings (those indicated with dashed arrows in \cref{fig:pblock}) of the 9 second- to fourth-row p-block elements calculated with 4c-SA-CASSCF, 4c-CASPT2, 4c-MR-CISD+Q, and MRPT2 and 3 to the experimental splittings. Flow parameters, $s$, are in units of $E_{\mathrm{h}}^{-2}$, and all results are reported in units of cm$^{-1}$, unless otherwise noted. ``Exp.'' stands for experimental splitting, and ``MAPE'' stands for mean absolute percentage error.}
\label{tab:hidden-spl}
\end{table}

\newpage

\section{Summary statistics over the main 15 spin-orbit splittings}
\begin{table}[!htb]
\begin{tabular}{cccccc}
\hline
\multicolumn{2}{c}{Method}           & MSE / cm$^{-1}$ & MAE / cm$^{-1}$ & MSPE / $\%$ & MAPE / $\%$ \\ \hline
\multicolumn{2}{c}{CASSCF}           & -16.130   & 21.115   & -2.744        & 7.804        \\ 
\multicolumn{2}{c}{CASPT2}           & -49.168   & 49.261   & -9.704        & 10.771       \\ \hline
\multirow{3}{*}{\shortstack[c]{DSRG-\\MRPT2}} & $s=0.5$  & -37.949   & 38.392   & -7.546        & 8.958        \\
                            & $s=0.35$ & -27.853   & 28.503   & -6.043        & 7.081        \\
                            & $s=0.2$  & -14.024   & 17.891   & -3.556        & 5.370        \\ \hline
\multirow{3}{*}{\shortstack[c]{DSRG-\\MRPT3}} & $s=0.5$  & -10.581   & 13.636   & -3.582        & 5.758        \\
                            & $s=0.35$ & -2.596   & 7.465   & -2.428        & 4.632        \\
                            & $s=0.2$  & 7.620   & 18.941   & -0.787        & 5.305        \\ \hline

\end{tabular}
\caption{The summary error statistics for the selected four-component methods over second- to fourth-row p-block elements.}
\label{tab:splitting_err_overall}
\end{table}

\clearpage

\section{Summary statistics broken down into groups}
\begin{table}[!htb]
\begin{tabular}{ccllllll}
\hline
\multicolumn{2}{c}{Method}                            & MAE  & 13     & 14     & 15     & 16      & 17     \\ \hline
\multicolumn{2}{c}{\multirow{2}{*}{CASSCF}}           & abs. & 52.523 & 16.831 & 11.692 & 15.827  & 8.702  \\
\multicolumn{2}{c}{}                                  & pct. & 14.647 & 7.511  & 13.011 & 1.945   & 1.903  \\ \hline
\multicolumn{2}{c}{\multirow{2}{*}{CASPT2}}           & abs. & 29.856 & 26.745 & 33.149 & 98.929  & 57.628 \\
\multicolumn{2}{c}{}                                  & pct. & 7.768  & 10.403 & 22.090 & 10.180  & 3.413  \\ \hline
\multirow{6}{*}{\shortstack[c]{DSRG-\\MRPT2}} & \multirow{2}{*}{$s=0.5$}  & abs. & 28.663 & 6.031  & 9.000  & 103.224 & 45.041 \\
                            &                         & pct. & 7.205  & 4.413  & 14.377 & 15.482  & 3.314  \\ \cline{2-8}
                            & \multirow{2}{*}{$s=0.35$} & abs. & 30.294 & 5.763  & 4.788  & 71.755  & 29.917 \\
                            &                         & pct. & 7.538  & 3.317  & 9.892  & 11.943  & 2.716  \\ \cline{2-8}
                            & \multirow{2}{*}{$s=0.2$}  & abs. & 34.017 & 6.053  & 3.954  & 32.046  & 13.385 \\
                            &                         & pct. & 8.399  & 3.929  & 5.062  & 7.015   & 2.443  \\ \hline
\multirow{6}{*}{\shortstack[c]{DSRG-\\MRPT3}} & \multirow{2}{*}{$s=0.5$}  & abs. & 11.525 & 7.638  & 2.363  & 29.064 & 17.591 \\
                            &                         & pct. & 4.155  & 5.441  & 11.554 & 6.391  & 1.250  \\ \cline{2-8}
                            & \multirow{2}{*}{$s=0.35$} & abs. & 13.084 & 6.101  & 2.924  & 8.589  & 6.627 \\
                            &                         & pct. & 4.614  & 4.273  & 1.270
  & 4.489  & 2.716  \\ \cline{2-8}
                            & \multirow{2}{*}{$s=0.2$}  & abs. & 17.266 & 3.785  & 9.087  & 30.749  & 33.818 \\
                            &                         & pct. & 5.687  & 2.600  & 9.726  & 5.788   & 2.722  \\ \hline

\end{tabular}
\caption{Summary error statistics for selected four-component methods across groups 13 through 17. The mean absolute errors (abs.) are in cm$^{-1}$, and the mean absolute percentage errors (pct.) are in percentage points.}
\label{tab:err_groups}
\end{table}
\clearpage

\section{Summary statistics broken down into rows}
\begin{table*}[!htb]
\footnotesize
\bgroup
\def\arraystretch{0.8}
\begin{tabular}{clllll}
\hline
\multicolumn{2}{c}{Method}           & MSE / cm$^{-1}$ & MAE / cm$^{-1}$ & MSPE / \% & MAPE / \% \\ \hline
\multicolumn{6}{c}{B-F}                                                                    \\ \hline
\multicolumn{2}{c}{CASSCF}           & -5.548    & 6.496    & -0.730        & 11.622       \\
\multicolumn{2}{c}{CASPT2}           & -6.889    & 7.167    & -4.503        & 7.702        \\
\multirow{3}{*}{\shortstack[c]{DSRG-\\MRPT2}} 
                            & $s=0.5$  & -12.771   & 13.295   & -12.258       & 15.451       \\
                            & $s=0.35$ & -11.796   & 12.144   & -9.343        & 11.462       \\
                            & $s=0.2$  & -9.812    & 10.339   & -4.532        & 7.742        \\ 
\multirow{3}{*}{\shortstack[c]{DSRG-\\MRPT3}} 
                            & $s=0.5$  & -6.721    & 7.125    & -8.160        & 10.623       \\
                            & $s=0.35$ & -6.564    & 6.876    & -5.272        & 7.170      \\
                            & $s=0.2$  & -6.137    & 6.541    & -2.119        & 5.763        \\ \hline
\multicolumn{6}{c}{Al-Cl}                                                                  \\ \hline
\multicolumn{2}{c}{CASSCF}           & -4.019    & 5.052    & -4.101        & 4.362        \\
\multicolumn{2}{c}{CASPT2}           & -14.369   & 14.369   & -10.980       & 10.980       \\
\multirow{3}{*}{\shortstack[c]{DSRG-\\MRPT2}} & $s=0.5$  & -12.146   & 12.950   & -3.630        & 4.673        \\
                            & $s=0.35$ & -5.983    & 8.395    & -3.631        & 4.627        \\
                            & $s=0.2$  & -0.308    & 5.774    & -3.304        & 4.238        \\ 
\multirow{3}{*}{\shortstack[c]{DSRG-\\MRPT3}} 
                            & $s=0.5$  & -2.181    & 4.413    & -1.425        & 4.320        \\
                            & $s=0.35$ & 1.820     & 4.199    & -2.014        & 5.037        \\
                            & $s=0.2$  & 7.387     & 10.685   & -1.877        & 6.340        \\ \hline

\multicolumn{6}{c}{Ga-Br}                                                                  \\ \hline
\multicolumn{2}{c}{CASSCF}           & -38.823   & 51.797   & -3.399        & 7.427        \\
\multicolumn{2}{c}{CASPT2}           & -126.248  & 126.248  & -13.630       & 13.630       \\
\multirow{3}{*}{\shortstack[c]{DSRG-\\MRPT2}} 
                            & $s=0.5$  & -88.930   & 88.930   & -6.751        & 6.751        \\
                            & $s=0.35$ & -64.971   & 64.971   & -5.154        & 5.154        \\
                            & $s=0.2$  & -31.952   & 37.560   & -2.832        & 4.130        \\ 
\multirow{3}{*}{\shortstack[c]{DSRG-\\MRPT3}} 
                            & $s=0.5$  & -22.842   & 29.371   & -1.160        & 2.332        \\
                            & $s=0.35$ & -3.043    & 11.320   & 0.000         & 1.688        \\
                            & $s=0.2$  & 21.609    & 39.597   & 1.633         & 3.811        \\ \hline

\end{tabular}
\egroup
\caption{Summary statistics of selected four-component methods for each row. Mean signed and average errors (MSE and MAE) and mean signed and average percentage errors (MSPE and MAPE) are reported.}
\label{tab:errors}
\end{table*}
\clearpage

\section{Results for calculations using DC coefficients}
We have experimented with using only the DC Hamiltonian for the 4c-CASSCF stage, using the resulting coefficients to transform the DCB atomic spinor integrals, which are then fed into 4c-DSRG-MRPT2. 
We can see that the performance is very similar to those using full DCB molecular spinor coefficients. 
This demonstrates that iterative reference relaxation can largely correct for inexact starting CI and MO coefficients, and can therefore be used to bypass expensive iterative DCB integral transformations in the SCF stages. 
This is further supported by the fact that, if the same MO coefficients used in the AO to MO transformation using DC atomic spinor integrals (``0.2 DCDC''), the resulting error distribution deteriorates significantly.

\begin{figure}[!hbt]
\centering
  \includegraphics[width=6.5in]{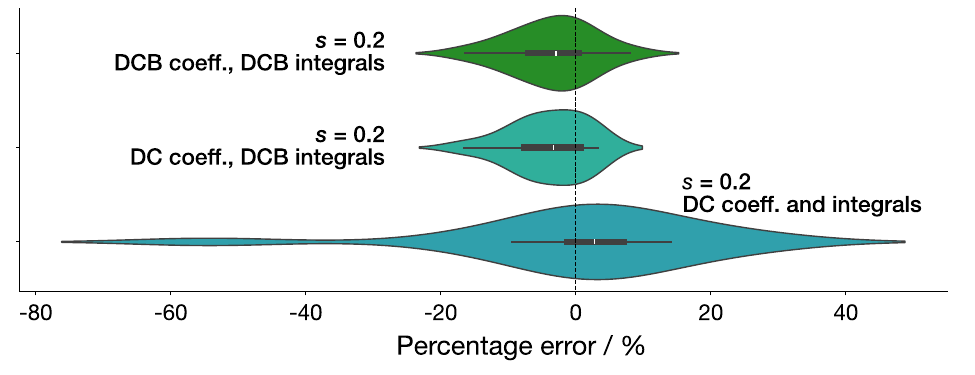}
  \caption{Distributions of the splitting errors for $s=0.2$ $E_{\mathrm{h}}^{-2}$, using DCB-CASSCF molecular spinor coefficients (green) or DC-CASSCF coefficients (teal) with DCB atomic spinor integrals, and using DC-CASSCF coefficients with DC atomic spinor integrals (blue)}
\end{figure}

\begin{table}[!htb]
\begin{threeparttable}
\begin{tabular}{lcccc}
\hhline{=====}
Splitting\hspace{50pt} & \hspace{15pt}Exp.\hspace{15pt}     & \hspace{15pt}0.2\hspace{15pt}  & \hspace{15pt}0.2 DC\hspace{15pt}  & \hspace{8pt}0.2 DCDC\hspace{8pt} \\ \hline
B $^2$P$_{\sfrac{1}{2}}\rightarrow^2$P$_{\sfrac{3}{2}}$  & 15.29   & 14.25   & 13.99   & 19.42    \\
C $^3$P$_{0}\rightarrow^3$P$_{1}$                        & 16.42   & 17.73   & 16.85   & 20.15    \\
N $^2$D$_{\sfrac{5}{2}}\rightarrow^2$D$_{\sfrac{3}{2}}$  & 8.71    & 8.56    & 8.33    & 3.98     \\
O $^3$P$_{2}\rightarrow^3$P$_{1}$                        & 158.27  & 132.32  & 132.05  & 133.39   \\
F $^2$P$_{\sfrac{3}{2}}\rightarrow^2$P$_{\sfrac{1}{2}}$  & 404.14  & 380.90  & 380.39  & 423.01   \\
Al $^2$P$_{\sfrac{1}{2}}\rightarrow^2$P$_{\sfrac{3}{2}}$ & 112.06  & 104.01  & 104.00  & 110.79   \\
Si $^3$P$_{0}\rightarrow^3$P$_{1}$                       & 77.12   & 77.78   & 78.27   & 82.35    \\
P $^2$D$_{\sfrac{3}{2}}\rightarrow^2$D$_{\sfrac{5}{2}}$  & 15.61   & 14.01   & 14.37   & 17.82    \\
S $^3$P$_{2}\rightarrow^3$P$_{1}$                        & 396.06  & 390.51  & 391.87  & 407.04   \\
Cl $^2$P$_{\sfrac{3}{2}}\rightarrow^2$P$_{\sfrac{1}{2}}$ & 882.35  & 895.35  & 895.35  & 932.38   \\
Ga $^2$P$_{\sfrac{1}{2}}\rightarrow^2$P$_{\sfrac{3}{2}}$ & 826.19  & 733.23  & 729.89  & 748.70   \\
Ge $^3$P$_{0}\rightarrow^3$P$_{1}$                       & 557.13  & 540.96  & 540.82  & 553.96   \\
As $^2$D$_{\sfrac{3}{2}}\rightarrow^2$D$_{\sfrac{5}{2}}$ & 322.10  & 332.20  & 332.72  & 346.24   \\
Se $^3$P$_{2}\rightarrow^3$P$_{1}$                       & 1989.50 & 1924.86 & 1924.60 & 1963.86  \\
Br $^2$P$_{\sfrac{3}{2}}\rightarrow^2$P$_{\sfrac{1}{2}}$ & 3685.24 & 3689.16 & 3689.57 & 3759.85  \\ \hline
MAE  &--&17.9&18.2&22.1\\
MAPE &--&5.4\%&5.2\%&11.7\%\\
\hhline{=====}
\end{tabular}
\end{threeparttable}
\caption{Comparison between the spin-orbit splittings of the 15 second- to fourth-row p-block elements calculated with MRPT2 with standard DCB Hamiltonian (`0.2'), MRPT2 with MO coefficients from a DC-CASSCF calculation transformed with DCB integrals (`0.2 DC'), and MRPT2 with MO coefficient from a DC-CASSCF calculation transformed with DC integrals (`0.2 DCDC') to the experimental splittings. All calculations employed a flow parameter, $s$, of 0.2 $E_{\mathrm{h}}^{-2}$, and all results are reported in units of cm$^{-1}$, unless otherwise noted. `Exp.' stands for experimental splitting, `MAE' stands for mean absolute error, and `MAPE' stands for mean absolute percentage error.}
\label{tab:dcdc}
\end{table}
\clearpage

\section{Results for using the augmented cc-pVTZ basis}
We also augmented other second row elements with a p function each with exponents determined by scaling the fluorine exponent by the ratio between the exponents of the tightest p functions in the original basis set of a given element and fluorine, resulting in exponents of 68.8243, 106.9535, 152.2273, and 196.9866 for boron, carbon, nitrogen, and oxygen respectively.

\begin{table}[!htp]
\scriptsize
\begin{tabular}{lccccccc}
\hhline{========}
\multirow{2}{*}{Splitting} &
\multirow{2}{*}{Exp.} & 
\multirow{2}{*}{CASSCF orig.} & 
\multirow{2}{*}{CASSCF aug.} & 
MRPT2 orig.& 
MRPT2 aug. & 
MRPT3 orig. & 
MRPT3 aug.\\[-7pt]
&&&&($s=0.24$)&($s=0.35$)&($s=0.24$)&($s=0.35$)\\ \hline
B $^2$P$_{\sfrac{1}{2}}\rightarrow^2$P$_{\sfrac{3}{2}}$ & 15.29    & 13.25     & 13.77    & 13.99    & 14.81    & 14.32    & 14.88    \\
C $^3$P$_{0}\rightarrow^3$P$_{1}$                       & 16.42    & 14.94     & 15.42    & 14.93    & 17.51    & 17.20    & 17.75    \\
C $^3$P$_1\rightarrow^3$P$_2$                           & 27.00    & 23.82     & 24.77    & 28.48    & 29.56    & 28.72    & 29.81    \\
N $^2$D$_{\sfrac{5}{2}}\rightarrow^2$D$_{\sfrac{3}{2}}$ & 8.71     & 11.09     & 11.01    & 9.41     & 8.02     & 7.89     & 7.73     \\
N $^4$S$_{\sfrac{3}{2}}\rightarrow^2$D$_{\sfrac{5}{2}}$ & 19224.46 & 22910.28  & 22914.03 & 20231.34 & 20235.14 & 19952.54 & 19956.74 \\
O $^3$P$_{2}\rightarrow^3$P$_{1}$                       & 158.27   & 153.23    & 157.34   & 145.35   & 133.37   & 138.92   & 142.57   \\
O $^3$P$_1\rightarrow^3$P$_0$                           & 68.71    & 66.99     & 68.95    & 54.49    & 56.12    & 58.32    & 60.07    \\
F $^2$P$_{\sfrac{3}{2}}\rightarrow^2$P$_{\sfrac{1}{2}}$ & 404.14   & 382.58    & 393.36   & 384.70   & 391.15   & 391.68   & 403.10   \\
\hhline{========}
\end{tabular}
\caption{Splittings in all 8 calculated spin-orbit splittings for second-row elements before and after augmenting the cc-pVTZ basis with a tight p function. Flow parameters are in units of $E_{\mathrm{h}}^{-2}$, `orig.' stands for `original basis', and `aug.' stands for `augmented basis'. All results are in cm$^{-1}$.}
\end{table}

\begin{table}[!htp]
\begin{tabular}{lccc}
\hhline{====}
\multirow{2}{*}{Splitting} &
\hspace{8pt}\multirow{2}{*}{CASSCF}\hspace{8pt} & 
\hspace{8pt}MRPT2\hspace{8pt} & 
\hspace{8pt}MRPT3\hspace{8pt}\\[-7pt]
&&($s=0.24$)&($s=0.35$)\\ \hline
B $^2$P$_{\sfrac{1}{2}}\rightarrow^2$P$_{\sfrac{3}{2}}$ & 3.40   & 5.95   & 3.78  \\
C $^3$P$_{0}\rightarrow^3$P$_{1}$                       & 2.94   & 16.73  & 3.14  \\
C $^3$P$_1\rightarrow^3$P$_2$                           & 3.53   & 4.36   & 3.69  \\
N $^2$D$_{\sfrac{5}{2}}\rightarrow^2$D$_{\sfrac{3}{2}}$ & -0.87  & -12.62 & -2.00 \\
N $^4$S$_{\sfrac{3}{2}}\rightarrow^2$D$_{\sfrac{5}{2}}$ & 0.02   & 0.02   & 0.02  \\
O $^3$P$_{2}\rightarrow^3$P$_{1}$                       & 2.59   & -7.61  & 2.74  \\
O $^3$P$_1\rightarrow^3$P$_0$                           & 2.85   & 2.36   & 3.12  \\
F $^2$P$_{\sfrac{3}{2}}\rightarrow^2$P$_{\sfrac{1}{2}}$ & 2.67   & 1.64   & 2.92  \\ \hline
Avg. improvement &2.14&1.35&2.18 \\ \hline
MAPE (orig.) & 11.44 & 8.74 & 7.64 \\
MAPE (aug.)  & 9.18  & 8.72 & 7.38 \\
\hhline{====}
\end{tabular}
\caption{Percentage error improvements in all 8 calculated spin-orbit splittings for second-row elements after augmenting the cc-pVTZ basis with a tight p function. A positive percentage signifies that the result after the augmentation is closer to the experimental splitting, and \textit{vice versa} for a negative percentage. Flow parameters are in units of $E_{\mathrm{h}}^{-2}$, ``Avg.'' stands for ``average'', ``MAPE'' stands for mean absolute percentage error, ``orig.'' stands for ``original basis'', and ``aug.'' stands for ``augmented basis''. All results are in percentage points.}
\label{tab:augment}
\end{table}

\clearpage
\bibliography{main}